\newif\ifsubmode
\submodefalse

\newif\ifprintfig
\printfigtrue

\newif\ifemulate
\emulatetrue

\ifsubmode
  \documentclass[12pt,preprint]{aastex}
  \received{}
  \accepted{}
  \journalid{}{}
  \articleid{}{}
\else
   \documentclass{emulateapj}
   \submitted{{\it To be submitted for publication in ApJL}}
\fi

\slugcomment{Draft \today}
\shortauthors{Hargis, Willman, and Peter}
\shorttitle{Milky Way Dwarfs in DES and LSST}

\begin{document}

\title{Too Many, Too Few, or Just Right? The Predicted Number and
  Distribution of Milky Way Dwarf Galaxies}

\author{Jonathan R. Hargis and Beth Willman} 

\affil{Department of Astronomy, Haverford College, 370 Lancaster Avenue, Haverford, PA 19041, USA; jhargis@haverford.edu}

\author{Annika H. G. Peter}

\affil{CCAPP and Department of Physics, The Ohio State University, 191 W. Woodruff Ave., Columbus, OH 43210, USA and
Department of Astronomy, The Ohio State University, 140 W. 18th Ave., Columbus, OH 43210, USA}

\begin{abstract}

We predict the spatial distribution and number of Milky Way dwarf galaxies to be discovered in the DES and LSST surveys, by completeness correcting the observed SDSS dwarf population.  We apply most massive in the past, earliest forming, and earliest infall toy models to a set of dark matter-only simulated Milky Way/M31 halo pairs from Exploring the Local Volume In Simulations (ELVIS).  The observed spatial distribution of Milky Way dwarfs in the LSST-era will discriminate between the earliest infall and other simplified models for how dwarf galaxies populate dark matter subhalos.  Inclusive of all toy models and simulations, at 90\% confidence we predict a total of  37--114 L $\gtrsim 10^3$L$_{\odot}$ dwarfs and 131--782  L $\lesssim 10^3$L$_{\odot}$ dwarfs within 300 kpc.  These numbers of L $\gtrsim 10^3$L$_{\odot}$  dwarfs are dramatically lower than previous predictions, owing primarily to our use of updated detection limits and the decreasing number of SDSS dwarfs discovered per sky area.  For an effective $r_{\rm limit}$ of 25.8 mag, we predict:  3--13 L $\gtrsim 10^3$L$_{\odot}$ and 9--99  L $\lesssim 10^3$L$_{\odot}$ dwarfs for DES, and 18--53 L $\gtrsim 10^3$L$_{\odot}$ and 53--307  L $\lesssim 10^3$L$_{\odot}$ dwarfs for LSST.   These enormous predicted ranges ensure a coming decade of near-field excitement with these next generation surveys.  
\end{abstract}

\section{Introduction}
\label{intro}

In a universe described by a $\Lambda$+cold dark matter cosmology ($\Lambda$CDM, \citealt{planck14}), the predicted number of dark matter subhalos far exceeds the observed number of dwarf galaxies orbiting Milky Way-like galaxies.  In the time since this discrepancy was dubbed `the missing satellites problem' \citep{kauffmann93a,klypin99a,moore99a}, more than a dozen previously unseen ultra-faint (L $\lesssim$ 50,000 L$_{\odot}$) Local Group dwarfs have been discovered, primarily in SDSS \citep[e.g.][and references therein]{willman10a} and PAndAS \citep[e.g.][]{martin13b,martin13a}.  However, it is not yet clear whether the observed number, spatial distribution, and masses of these discoveries are consistent with CDM-based expectations \citep{weinberg13a}.

A common approach to cosmological interpretation of the MW's dwarf population is to predict the number \citep[e.g.][]{tollerud08a} and spatial distribution \citep[e.g.][]{willman04a,maccio10a,yniguez14a} expected in surveys, such as the Dark Energy Survey (DES, \citealt{DES,rossetto11a}) and the Large Synoptic Survey Telescope (LSST, \citealt{LSST}).  Such predictions are typically either 
 i) based on an N-body simulation+semi-analytic galaxy formation model \citep{maccio10a,bovill11b,font11a}, or ii) a completeness correction of the known number of MW dwarfs, using a physically motivated spatial distribution  \citep{koposov08a,tollerud08a,walsh09a}.  All past predictions for ultra-faint dwarf galaxies in DES and LSST have been subject to one or more limitations: i) the use of only one average model or a single simulation to predict the spatial distribution of ultra-faint dwarfs, ii) the use of isolated (rather than paired MW/M31) galaxy simulations, iii) the inclusion of ``hyperfaint" (L $\lesssim$ 10$^{\rm 3}$ L$_{\odot}$) dwarfs such as Segue 1 in the same predictions as more luminous dwarfs, or iv) the application of rigid magnitude limits for DES and LSST. 
 
 In this paper, we take steps to overcome these limitations using the Exploring the Local Volume In Simulations \citep[ELVIS,][]{garrison-kimmel2014}   suite of N-body simulations, plus physically motivated toy models for populating subhalos with galaxies. We predict a spatial distribution of MW dwarf galaxies and correct the MW dwarf galaxy count for completeness.   In Section~\ref{sims} we summarize our toy models.  In Section~\ref{sec:spatial} we discuss the predicted spatial distributions of dwarfs, and in Section~\ref{sec:results} we present our predicted numbers. We summarize our expectations for DES and LSST in Section~\ref{conclusion}.

\section{Dwarf Galaxy Toy Models}
\label{sims}

Owing to observational bias, the underlying spatial distribution of the MW's dwarf galaxy population is unknown.  To statistically correct the observed number of MW dwarfs, we use spatial distributions of dark matter subhalos in the ELVIS simulations.  ELVIS includes a dozen dark matter-only, cosmological zoom-in simulations of MW and Andromeda pairs (see \citealt{garrison-kimmel2014} for details).  We consider the galaxy in each simulated pair with the lower virial mass to be the MW analog.  The particle mass in the high-resolution regions of the fiducial simulations is  $m_p = 1.9\times 10^5 M_\odot$.  The simulation suite includes isolated simulations of each of the MW and Andromeda analogs at the same resolution,  with three simulated at a higher mass resolution of $m_p = 2.35\times 10^4 M_\odot$.  The fiducial subhalo catalogs are complete to $V_{max} = 8\hbox{ km s}^{-1}$ and $V_{peak} = 12\hbox{ km s}^{-1}$. 

To generate statistical descriptions for the spatial distribution of MW dwarfs, we implement three physically motivated toy models for which subhalos of the ELVIS MW analogs host dwarf galaxies - most massive in the past, earliest forming, and earliest infall.   The first two models have been implemented in a number of past studies \cite[e.g.,][]{bullock00a,strigari07a, kravtsov10a}.  For all models we exclude any subhalos with $V_{peak}>25\hbox{ km s}^{-1}$, as their deeper potential wells are likely to host the more luminous ``classical'' dwarfs:

$\bullet$ {\bf Massive in the past} ($V_{peak} > 12\hbox{ km s}^{-1}$):  Subhalos with deeper potential wells are more likely to retain and cool the gas fuel necessary for star formation.   $V_{peak}$ is the historical peak of a subhalo's circular velocity curve and provides a measure of potential well depth.    We tested threshold values for $V_{peak}$ of 12 $\hbox{ km s}^{-1}$ and higher, and found an average of $\sim100$ subhalos per MW-massed primary for $V_{peak} > 15\hbox{ km s}^{-1}$ and $\sim200$ for $V_{peak} > 12\hbox{ km s}^{-1}$.  Although the typical number of subhalos in the higher threshold model is a better match to the predicted number of ultra-faint dwarfs, the spatial distribution of the higher threshold model is similar.  We therefore implement the $V_{peak} >$ 12 km s$^{\rm -1}$  cut to improve statistics.  

$\bullet$ {\bf Formed before reionization} ($z > 8$ and $N_{\rm part} \geq 32$):  Another hypothesis is that the ultra-faint dwarf galaxies are ``fossils" that produced the bulk of their stars prior to reionization \citep{bovill11b}.  This hypothesis is supported by photometric and spectroscopic studies of ultra-faint dwarfs, which show that these galaxies formed their stars within $\sim 1$ Gyr of each other at early times \citep{sand2010,brown2012,kirby2013,frebel2014}.  We follow the  \citet{bovill11b} definition of reionization fossils as galaxies residing in halos with $V_{peak} < 20\hbox{ km s}^{-1}$, and which had a resolved progenitor in the ELVIS paired simulations at $z=8$.  These thresholds result in $\sim 170$ subhalos per MW analog.  

$\bullet$ {\bf Earliest Infall} ($z_{peak} \geq 3$, i.e. $t_{\rm infall} \gtrsim 11.5$ Gyr ago, and $V_{max} > 8\hbox{ km s}^{-1}$):  An alternative model to explain the truncated star formation histories of ultra-faint dwarfs is that early infall into the MW's halo resulted in massive gas stripping \citep[e.g.][]{weisz14a}.  Many of the MW's ultra-faint dwarfs show spatial or kinematic hints of tidal disturbance \citep{munoz2010,willman2011,sand12a}, which might be expected among dwarfs that have been orbiting within the MW's potential for the longest time, and with (on average) relatively smaller orbital pericenters.  Because subhalos typically reach  $V_{peak}$ just before infall onto a larger halo \citep{behroozi2014}, we adopt the redshift of peak circular velocity as our estimate of the infall time.  To avoid overestimating the Poisson contribution to our uncertainties, for this model we exclude the four ELVIS pairs with resulting subhalo counts less than the number of $\gtrsim 10^3$L$_{\odot}$ dwarfs expected from an area-only correction ($\sim 34$; Zeus \& Hera, Sonny \& Cher, Hall \& Oates, Thelma \& Louise).  This model yields $\sim 56$ subhalos per MW analog.

\subsection{Numerical Testing}

Figure \ref{radial_profiles} compares the cumulative radial profile of each toy model applied to: the 12 MW analogs from the paired fiducial simulations, the corresponding 12 isolated fiducial simulations, the 3 high-res isolated simulations, and the 3 corresponding fiducial isolated simulations.  

These comparisons demonstrate that the radial distributions of subhalos in the paired and isolated simulations are indistinguishable, and the high-res and fiducial-res simulations are statistically the same when comparing the same three simulated halos.  The range of radial profiles from the 12 paired MW analogs are also shown, with the earliest infall model showing a large halo-to-halo variation.  We will further discuss these radial distributions in \S\ref{sec:spatial}.
  
A mild caveat for the ELVIS-based radial distributions is that baryonic physics can decrease the survivability of subhalos with small orbital pericenters.  For example, tidal shocking of satellites by the MW's disk can accelerate mass loss \citep{donghia2010}.  Supernova  feedback may yield cored density profiles in the most luminous dwarf satellites, making them more vulnerable to destruction \citep{zolotov2012,brooks2013}.  The former issue is more relevant to the ultra-faint dwarfs than the latter \citep{governato2012}.  However, both effects should drive the true surviving subhalo population to be less concentrated than the ELVIS toy models. Thus our predictions may be lower limits to the expected satellites counts in future surveys.
  
\section{Predicted Spatial Distribution of Dwarfs}\label{sec:spatial}

\begin{figure}
\plotone{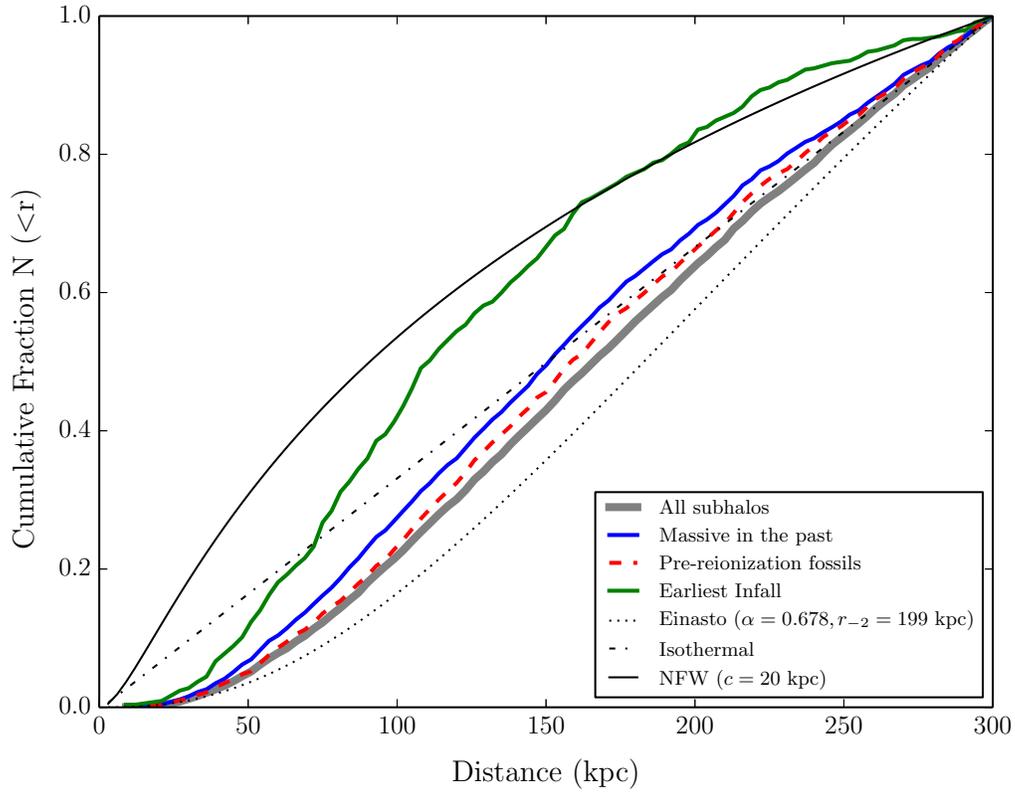}
\caption{Comparison of the mean cumulative radial distributions of subhalos for the three toy models applied to the MW analog in the 12 ELVIS paired simulations.  Overplotted are three common analytic descriptions for dark matter subhalos: Einasto, isothermal, and NFW.  The Einasto model matches the radial distribution of subhalos in the Aquarius simulations \citep{springel2008}. The NFW concentration is consistent with studies of the MW halo radial velocity dispersion profile \citep{battaglia2005} and rotation curve \citep{deason2012a}.
\label{radial_profiles2}}
\end{figure}

A complete census of dwarfs out to the MW's virial radius should provide the statistics needed to use the observed radial distribution of dwarfs to discriminate between some toy models for how dwarfs populate subhalos \citep[e.g.][]{gao2004a, willman04a, rocha2012, wang2013a}. Figure~\ref{radial_profiles} shows that the earliest infall subhalos are significantly more centrally concentrated than the most massive and earliest forming subhalos, which have very similar spatial distributions to that of all subhalos resolved in ELVIS.  Figure \ref{radial_profiles2} compares the radial distributions of each toy model to Einasto, isothermal, and NFW radial profiles.  All three toy models have profiles between Einasto and NFW, with the early infall model best described by an NFW profile and the most massive and earliest forming models best described by an isothermal.  While the mean differences in these toy model radial distributions cause the mean differences in the number of dwarfs we predict for each model, the halo-to-halo differences within each toy model (Figure  \ref{radial_profiles}) contribute to the statistical uncertainty in each of our model predictions (see \S\ref{sec:results}). 

The lack of MW dwarf discoveries since the SDSS Data Release 6 (DR6, 9100 deg$^2$, 12 dwarfs), has recently re-invigorated the discussion of the population's azimuthal anisotropy.  Inspired by this observational result, we explored $\Lambda$CDM expectations by generating 100 mock survey pointings for each of five survey areas ($1000-14500\hbox{ deg}^2$) in both the paired ELVIS MW analogs and a set of azimuthally uniform subhalo distributions (``Poisson" expectations).  The survey-to-survey variation in subhalo counts (for the median of all simulations) is marginally consistent with Poisson expectations on survey scales larger than $\sim$ 10,000 deg$^2$.  On smaller survey scales, however, Poisson fluctuations are a lower-limit to the expected survey-to-survey variations.  Most ELVIS simulations ($\sim$ 9 of the 12) show survey-to-survey variations of up to twice those of Poisson expectations, although usually only $\sim$25-50\% more variation, underscoring the importance of accounting for azimuthal variations when making predictions (see \S\ref{sec:results}).  We note that the lack of dwarfs in SDSS DR8 is actually consistent with Poisson expectations when only considering the 2900 deg$^2$ of new imaging area at $|b| > 25$ deg.  There is a significant observational bias against finding dwarfs within 25 degrees of the MW's disk \citep{walsh09a}.

To look specifically for evidence of a systematic alignment of subhalos along the MW-M31 axis, we generated smoothed spatial maps of the resolved subhalo distributions (stacked, medians, and individual), rotated to place the simulated M31 analog at the same Galactic coordinates as M31.  Although the stacked map showed a suggestive trend, we did not find a statistically significant global trend in the alignment of the subhalo distribution relative to M31.  Examining the individual maps showed that only three paired simulations displayed an overdensity of subhalos in the direction of the M31 analog (Burr \& Hamilton, Scylla \& Charybdis, and Kek \& Kauket).

\section{Predicted Numbers of Dwarfs in DES and LSST}
\label{sec:results}

We make our predictions using a method similar to \citet{tollerud08a}.  Our approach has two steps: i) completeness correcting the observed MW dwarfs to the total number of expected within 300 kpc within a randomly placed mock-SDSS footprint, then ii) scaling that result to a randomly placed mock DES (5000 deg$^2$) or LSST (20,000 deg$^2$) survey with some magnitude limit.  For each toy model and each fiducial survey magnitude limit, we simulate 100 mock-SDSS + random survey pointings in each MW analog in the 12 paired ELVIS simulations.  

We adopt the 12,000 deg$^2$ above $|b| >$ 25 deg for the SDSS survey area because it is the portion of the SDSS DR8 footprint with uniform detection limits \citep{walsh09a} and it includes all 14 dwarfs discovered in SDSS. Our correction only includes these 14 dwarfs, because we assume that dwarfs similar to those previously known would have already been detected at $|b| >$ 25 deg \citep{willman10a}.  This includes 10 L $\gtrsim 10^3$ L$_{\odot}$ dwarfs (Hercules, Bo\"{o}tes I, Leo IV and V, Pisces II, Canes Venatici I and II, UMa I and II, Coma Berenices) and four  ``hyperfaint" L $\lesssim 10^3$ L$_{\odot}$ dwarfs (Bo\"{o}tes II, Willman 1, and Segue 1 and 2).   The boundary at L $\sim 10^3$ L$_{\odot}$ loosely corresponds to those objects that can only be discovered by main sequence turnoff stars and fainter (too few red giant branch stars).  

For each dwarf, the corrected number within the SDSS footprint is 1/(fraction of toy model subhalos within the maximum detection distance).  We calculate each dwarf's maximum detection distance, $d_{\rm max, SDSS}$, using the $90\%$ SDSS detection efficiency function given by \citet{walsh09a}.    This fraction is normalized to unity for $d_{\rm max, SDSS}$=300 kpc. Only CVn I has $d_{\rm max, SDSS} > 300$ kpc, resulting in no completeness correction.  The values of $d_{\rm max, SDSS}$ for CVn II, Psc II, and Leo V are smaller than their observed distances, so we adopt $d_{\rm observed}$ as their $d_{\rm max, SDSS}$'s and perform an additional efficiency correction based on an estimated integrated detection efficiency within $d_{\rm observed}$ from \citet[][$\epsilon$ = 1.0, 0.5, and 0.85 respectively]{walsh09a}.  Objects like the other 11 SDSS dwarfs were detected with 100\% efficiency with the \citet{walsh09a} algorithm, unlike the \citet{koposov08a} algorithm--the primary source of the difference between our and the \citet{tollerud08a} results.

To scale the corrected numbers within the SDSS footprint to the expected numbers in each mock DES or LSST survey, we account for both survey area and point-source detection limit.  Rather than scaling directly by relative survey area, we scale by the ratio of the number of subhalos within $d_{\rm max, survey}$ of a mock survey area to the number within $d_{\rm max, survey}$ of a mock SDSS.   This captures the azimuthal anisotropy in the ELVIS simulations, allowing us to directly incorporate the effect into our uncertainties.  We naively assume that completeness distances scale like the flux depth of each survey ($d_{\rm max,survey}$ = $d_{\rm max,SDSS}$ $\times$ 10$^{0.2 (r_{\rm lim, survey} - r_{\rm lim, SDSS})}$), given $r_{\rm lim, SDSS} = 22.0$ mag.  In light of the challenges separating resolved stars from unresolved galaxies at faint apparent magnitudes, we consider $23.3 < r_{\rm lim, survey}< 25.8$ for both DES and LSST. 

We apply a slightly different method to the hyperfaint dwarfs, because the numbers of subhalos within their $d_{\rm max, SDSS}$ ($\lesssim 50\hbox{ kpc}$) are too small to provide robust mock survey results.   In some toy models, several MW analogs have no subhalos within the $d_{\rm max}$ of Seg 1 ($\sim 30$ kpc).  We therefore used azimuthally-averaged radial distributions to predict an average number for each simulation and used the mock survey approach to estimate the 10/90 percent confidence intervals.  We ignored any random mock survey pointing without a Segue 1-like subhalo.  The resulting predictions do not properly capture halo-to-halo and spatial anisotropy uncertainties, but they provide reasonable limits on the uncertainty in the predicted numbers.

\begin{figure}
\centering
\includegraphics[scale=.55,angle=0,clip=true]{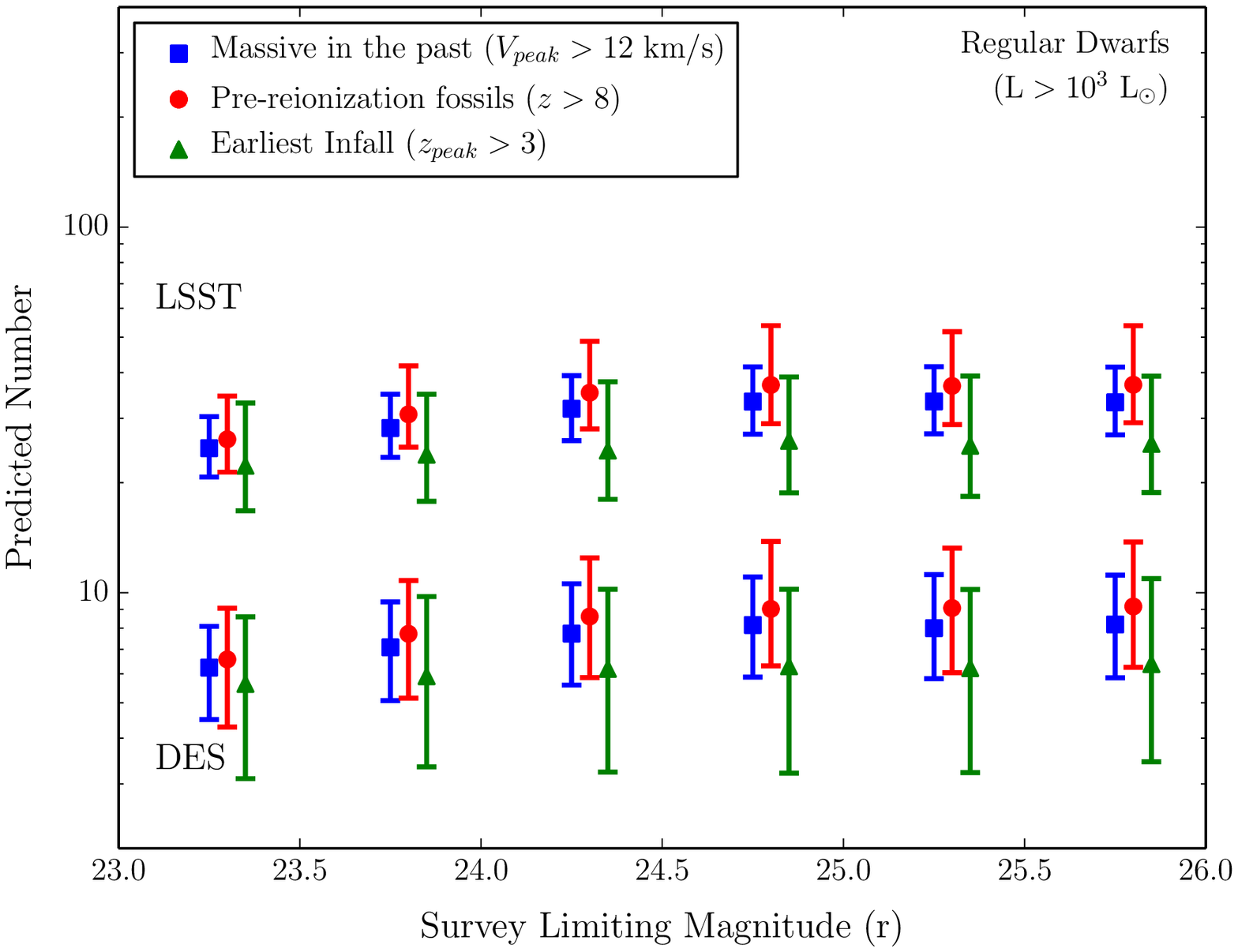}\\
\includegraphics[scale=.55,angle=0,clip=true]{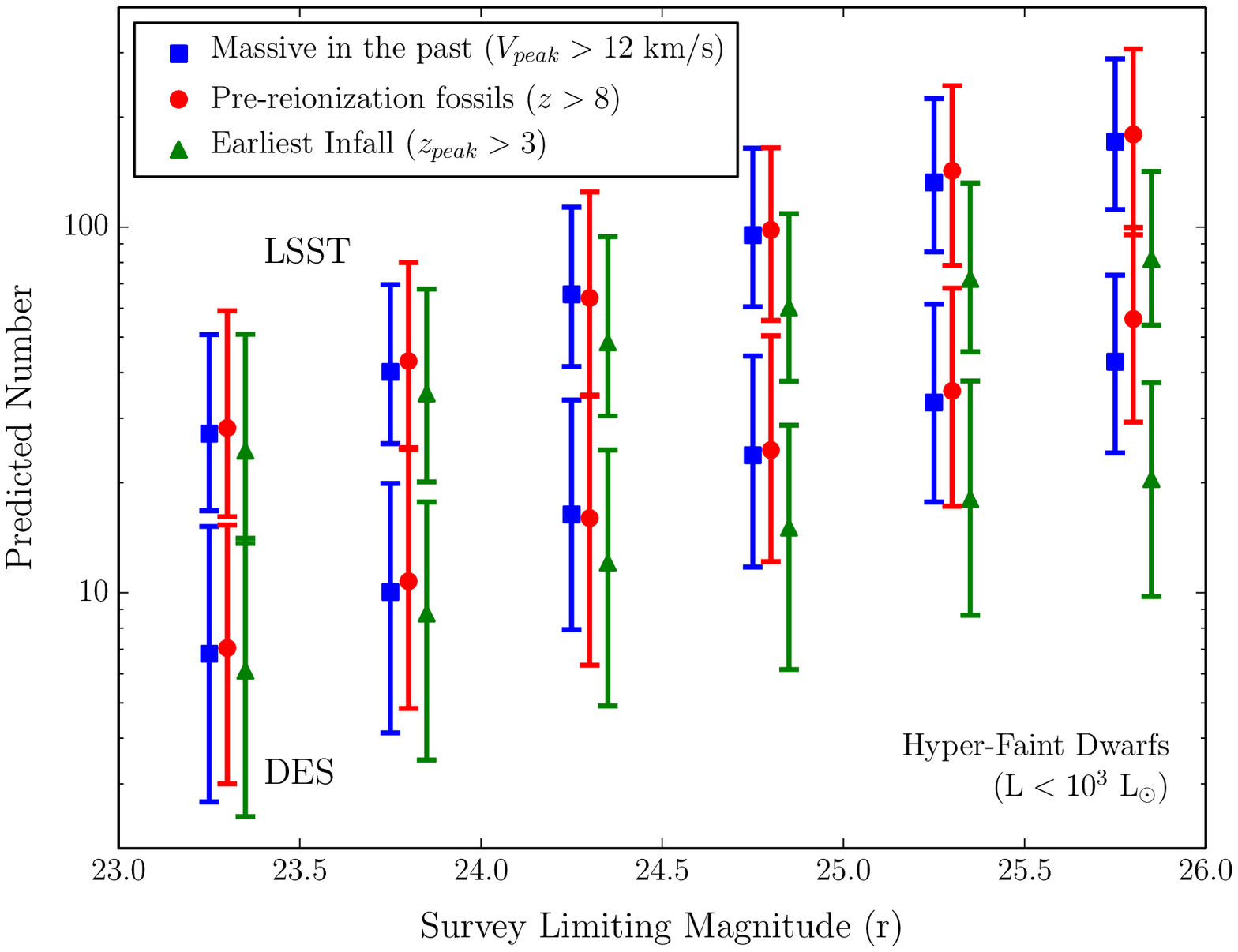}

\caption{\footnotesize Predicted number of ultra-faint dwarfs for each of the three toy models as a function of
  survey $r$ band limiting magnitude for LSST and DES.  The results for the brighter and fainter subsets of the ultra-faints are shown in the top and bottom panels, respectively. 
The error bars show the 10/90 percent confidence
  intervals as described in $\S\ref{sec:results}$.   
  \label{survey_summary}}
\end{figure}

The top and bottom panels of Figure~\ref{survey_summary} show the predicted numbers of  L $\gtrsim 10^3$ L$_{\odot}$ and L $\lesssim 10^3$ L$_{\odot}$ dwarfs as a function of $r$-magnitude depth.   We adopted the median and 10/90 percent confidence intervals from the 1200 mock surveys as our estimated number and uncertainty.  The uncertainty on each number reflects both halo-to-halo and survey-to-survey (radial and azimuthal) variations. 

\section{Discussion}
\label{conclusion}

Using our approach of correcting the known population of SDSS dwarfs, we find that the use of paired versus isolated simulations does not yield systematically different predictions for the MW's dwarf population.  Although there is statistically significant super-Poisson azimuthal anisotropy in the toy model subhalo distributions,  this anisotropy appears neither extreme nor systematically aligned with M31.

We predict vastly different numbers of regular (L $\gtrsim 10^3$ L$_{\odot}$) versus hyperfaint (L $\lesssim 10^3$ L$_{\odot}$) dwarfs (see Tables~\ref{tab:table1} and ~\ref{tab:table2}). Spanning all toy models, at 90\% confidence, and assuming $r_{\rm limit}$= 25.8 mag: 3 -- 13 regular vs. 9--99 hyperfaints should be discovered in DES and 18 -- 53 regular vs. 53--307 hyperfaints should be discovered in LSST.  Over the entire sky and within 300 kpc, we predict $\sim 37-114$ regular and $\sim 131-782$ hyperfaint dwarfs with 90\% confidence (see Table~\ref{tab:table2}).  Owing to the severely limited numbers of subhalos within $d_{\rm max}$ for the hyperfaint dwarfs, those predicted numbers are less robust and should be interpreted lightly in future observational studies.  We emphasize that these numbers assume no future bias against discovering dwarfs at low Galactic latitude.  Any interpretation of future observational studies must also cautiously account for this bias.

The predicted number of L $\gtrsim 10^3$ L$_{\odot}$ dwarfs is strikingly lower than previous predictions \citep[e.g.][]{tollerud08a} due to our use of updated detection limits, the decreased rate of SDSS dwarfs discovered per sky area, our toy models, and a 300 kpc distance limit.  The lower limit on the predicted number of L $\gtrsim 10^3$ L$_{\odot}$ is similar to a simple area correction, because a mock-SDSS can include an overdensity of subhalos and because the radial distribution of the early infall model of some MW analogs is very centrally concentrated.   The discovery of only a few dwarfs in DES imaging (for a shallow survey limit of $r\sim23.8$) could be consistent with our lower limits.

\begin{figure}
\plotone{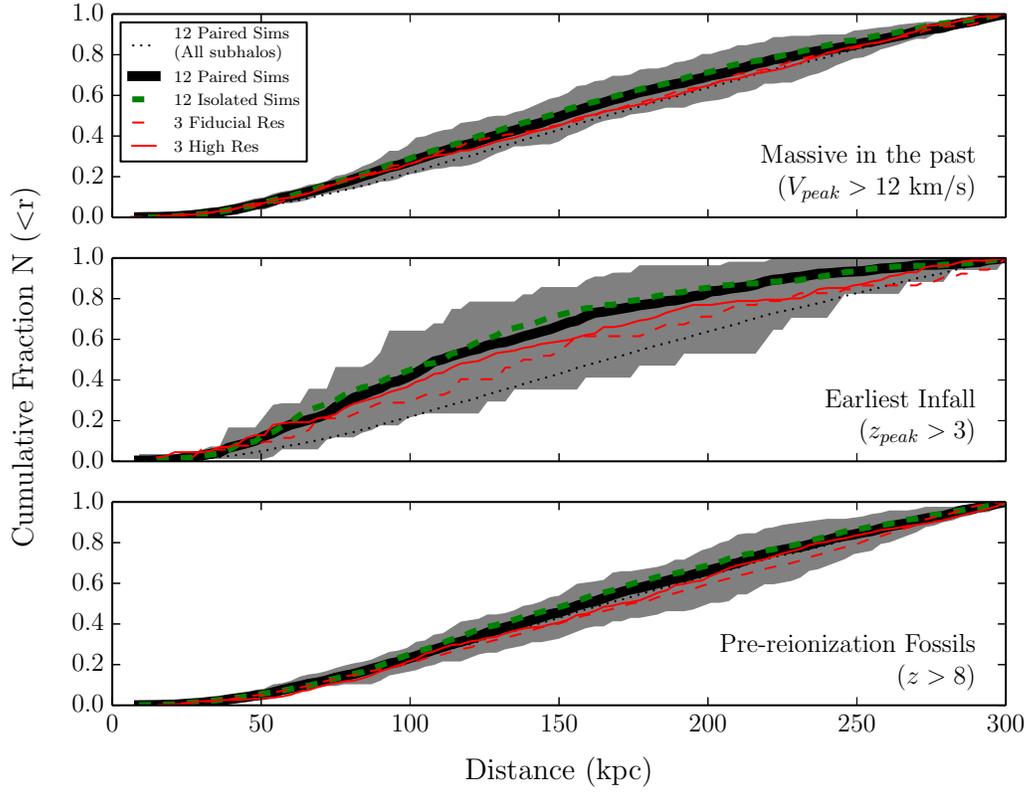}
\caption{Comparison of the cumulative radial distribution of subhalos for the three toy models: massive in the past (top panel); earliest infall (middle panel); pre-reionization fossils (bottom panel).  Shown are the mean profiles for the 12 paired  (\textit{solid black}), 12 isolated  (\textit{dashed green}), and three isolated fiducial  ({\it dashed red line}) and high ({\it solid red line}) resolution simulations.  The mean of the 12 paired fiducial resolution simulations is shown for comparison (\textit{dotted line}).  The grey region shows the simulation-to-simulation scatter from the 12 paired simulations. 
\label{radial_profiles}}
\end{figure}

The predicted total numbers of dwarfs are not significantly different between the most massive in the past or the pre-reionization models, but the earliest infall model systematically predicts lower total numbers of dwarfs by a factor of $\sim$2.  The discoveries of small numbers of dwarfs in future surveys would provide support for an early infall, or other centrally concentrated, model.  However, such hypotheses may be more sensitively tested by the observed radial distribution of dwarfs (e.g. Section~\ref{sec:spatial}, Figure~\ref{radial_profiles}).

Table~\ref{tab:table1} summarizes our DES and LSST predictions for two effective survey depths, $r_{\rm lim}$ = 23.8 and 25.8 mag.  Both of these are shallower than the expected surveys' 5$\sigma$ point source detection limits.  The shallower limit is a pessimistic estimate of the effective survey depth possible for studies of low-surface brightness objects if color-based methods to separate stars from unresolved galaxies are not successful, while the deeper limit may be achievable with color-based, probabilistic star-galaxy separation \citep{fadely12a}. 

Within their footprints, and at high Galactic latitudes, DES and LSST should easily recover the full population of MW dwarfs similar to those known with L $\gtrsim 10^3$ L$_{\odot}$.  For LSST, $r\sim 24$ corresponds to the expected single-visit imaging depth ($\sim0.5$ mag brighter than the $5\sigma$ point-source detection threshold) and will be sufficient to discover L $\gtrsim 10^3$ L$_{\odot}$ dwarfs.  Increased imaging depths will primarily yield the detection of increasing numbers of hyperfaint dwarfs.  Given the enormous range in our predicted numbers, the answer to the question, ``Is there a missing satellites problem with CDM?" is likely to be ``No" in the era of DES and LSST.     

\acknowledgments 

BW and JH were supported by an NSF Faculty Early Career Development (CAREER) award (AST-1151462).  This work was also supported in part by National Science Foundation Grant No. PHYS-1066293 and the hospitality of the Aspen Center for Physics.  We acknowledge Shea Garrison-Kimmel for helpful conversations and for sharing the ELVIS data.  We thank Tim Beers for inspiring our use of the term ``hyperfaint" dwarf.  We also acknowledge useful conversations with Mike Boylan-Kolchin, Andrew Wetzel, Erik Tollerud, and Alis Deason.



\begin{center}
\begin{deluxetable}{lcc}
\tablecolumns{3}
\tablewidth{20pc}
\tablecaption{Predicted Number of Dwarf Galaxies for LSST and DES\label{tab:table1}}
\tablehead{
\colhead {} & \colhead{DES ($\pm$ 10/90)} & \colhead{LSST ($\pm$ 10/90)} 
}
\startdata
$L > 10^3 L_\odot$, $r_{lim} = 23.8$ &  & \\
Massive in the past & $7^{+2}_{-2}$  & $28^{+6}_{-5}$\\
Pre-reionization Fossils & $7^{+3}_{-2}$& $30^{+11}_{-5}$\\
Earliest Infall & $5^{+4}_{-2}$ & $23^{+11}_{-6}$ \\
\hline
$L < 10^3 L_\odot$, $r_{lim} = 23.8$ &   & \\
Massive in the past &  $10^{+9}_{-6}$ & $40^{+29}_{-15}$\\
Pre-reionization Fossils & $10^{+14}_{-6}$ & $43^{+36}_{-19}$\\
Earliest Infall & $8^{+9}_{-5}$ & $35^{+32}_{-15}$\\
\hline
$L > 10^3 L_\odot$, $r_{lim} = 25.8$ &   & \\
Massive in the past & $8^{+3}_{-3}$  & $33^{+8}_{-6}$\\
Pre-reionization Fossils & $9^{+4}_{-3}$ & $37^{+16}_{-8}$\\
Earliest Infall & $6^{+4}_{-3}$ & $25^{+14}_{-7}$\\
\hline
$L < 10^3 L_\odot$, $r_{lim} = 25.8$ &   & \\
Massive in the past & $42^{+31}_{-18}$ & $171^{+117}_{-60}$\\
Pre-reionization Fossils & $56^{+43}_{-27}$ &$179^{+128}_{-84}$ \\
Earliest Infall & $20^{+17}_{-11}$  &$81^{+60}_{-28}$
\enddata
\end{deluxetable}
\end{center}

\begin{center}
\begin{deluxetable}{lc}
\tablecolumns{2}
\tablewidth{15pc}
\tablecaption{Predicted Number of Dwarf Galaxies within $d=300$ kpc \label{tab:table2}}
\footnotetext{These numbers do not include objects like the ``classical" dwarf galaxies.}
\tablehead{
\colhead {} & \colhead{All Sky ($\pm$ 10/90)}  
}
\startdata
$L > 10^3 L_\odot$ &   \\
Massive in the past & $69^{+19}_{-14}$  \\
Pre-reionization Fossils & $78^{+36}_{-21}$ \\
Earliest Infall &  $53^{+30}_{-16}$ \\
\hline
$L < 10^3 L_\odot$  &    \\
Massive in the past &  $477^{+305}_{-185}$ \\
Pre-reionization Fossils & $485^{+277}_{-246}$\\
Earliest Infall & $197^{+145}_{-66}$
\enddata
\end{deluxetable}
\end{center}


\end{document}